\documentclass[version = preprint]{iacrcc}

\license{CC-by}

\usepackage{algorithm, algorithmic, amsmath, graphicx} 
\usepackage{tabularx, soul, color, booktabs, url, multirow}

\title[running  = {DL-Assisted Improved DFAs on Lightweight Stream Ciphers},
      ]{Deep Learning-Assisted Improved Differential Fault Attacks on Lightweight Stream Ciphers}

\addauthor[orcid    = {0009-0001-4713-2663},
           inst     = {1},
           email    = {cys2209667@xmu.edu.my},
           surname  = {Lim}
          ]{Kok Ping Lim}

\addauthor[orcid    = {0009-0001-2164-242X},
           inst     = {1},
           email    = {cme2209122@xmu.edu.my},
           surname  = {Jia}
          ]{Dongyang Jia}

\addauthor[orcid    = {0000-0003-1395-4623},
           inst     = {1},
           email    = {iftekhar.salam@xmu.edu.my},
           surname  = {Salam}
          ]{Iftekhar Salam}

\addaffiliation[ror     = 0331wa828,
                street  = {},
                city    = {Sepang 43900},
                postcode= {},
                country = {Malaysia}
               ]{School of Computing and Data Science, \\Xiamen University Malaysia}

\makeatletter
\let\thecicdate\@empty
\makeatother

\begin{document}

\maketitle


\keywords[Differential Fault Attack, MLP, ACORN, MORUS, ATOM]{Differential Fault Attack, MLP, ACORN, MORUS, ATOM}

\begin{abstract}
  Lightweight cryptographic primitives are widely deployed in resource-constrained environments, particularly in Internet of Things (IoT) devices. Due to their public accessibility, these devices are vulnerable to physical attacks, especially fault attacks. Recently, deep learning–based cryptanalytic techniques have demonstrated promising results; however, their application to fault attacks remains limited, particularly for stream ciphers. In this work, we investigate the feasibility of deep learning assisted differential fault attacks on three lightweight stream ciphers, namely ACORNv3, MORUSv2, and ATOM, under a relaxed fault model in which a single-bit bit-flipping fault is injected at an unknown location. We develop and train multilayer perceptron (MLP) models to identify the fault locations. Experimental results show that the trained models achieve high identification accuracies of 0.999880, 0.999231, and 0.823568 for ACORNv3, MORUSv2 and ATOM, respectively, and outperform traditional signature-based methods. For the secret recovery process, we introduce a threshold-based method to optimize the number of fault injections required to recover the secret information. The results show that the initial state of ACORN can be recovered with 21 to 34 faults, while MORUS requires 213 to 248 faults, with at most 6 bits of guessing. Both attacks reduce the attack complexity compared to existing works. For ATOM, the results show that it possesses a higher security margin, as the majority of state bits in the Nonlinear Feedback Shift Register (NFSR) can only be recovered under a precise control model. To the best of our knowledge, this work provides the first experimental results of differential fault attacks on ATOM.
\end{abstract}

\begin{textabstract}
  For the final version of your paper, you will need a text-only abstract. Do
  not use LaTeX macros inside this abstract.
\end{textabstract}

\section{Introduction}
Over the past two decades, various cryptographic competitions, such as eSTREAM, the Competition for Authenticated Encryption: Security, Applicability, and Robustness (CAESAR), and the National Institute of Standards and Technology (NIST) Lightweight Cryptography Project, were initiated for the design and standardization of lightweight cryptographic primitives. These primitives ensure secure communication for resource-constrained devices, e.g., Internet of Things (IoT) devices, as traditional ciphers are intended to operate in resource-rich environments. Many cryptanalysis techniques, such as cube attacks, side channel attacks, and fault attacks are applied to the lightweight ciphers. Among these attacks, fault attacks have gained significant attention in recent years due to their effectiveness against physical hardware and the ongoing advancement of fault injection techniques. Since IoT devices are commonly deployed in publicly accessible areas, adversaries often have physical access to them, making these devices viable targets for fault attacks \cite{silvaetal2025}.

Fault attack was first introduced by Boneh et al. \cite{bonehetal1997} and targets physical hardware rather than the cryptographic algorithm itself. They applied the attack to RSA and Rabin signature schemes by introducing hardware faults to factor the large prime modulus $N$. In the same year, Biham and Shamir \cite{bihamshamir1997} implemented a variant of fault attack, differential fault attack (DFA), and applied it against the Data Encryption Standard (DES) to recover the 56-bit secret key. The attack works by inducing an error (fault) during the operation of the cipher to obtain the faulty outputs. By analyzing the differences between fault-free and faulty outputs, some secret information of the cipher, such as the internal state or the secret key can be recovered. Hoch and Shamir \cite{hochshamir2004} further extended DFA to stream ciphers such as RC4, LILI-128, and SOBER-t32.

Following these foundational works, differential fault attacks were further extended and refined, and subsequently applied to evaluate the security of a wide range of cryptographic primitives. Orumiehchiha et al. \cite{Orumiehchihaetal2020} applied the first DFA against WG-8 stream cipher to recover the initial state and secret key of the cipher using 6 faults with a time complexity of $2^{24}$. Ma, Tian and Qi \cite{maetal2021} applied DFA to the LIZARD stream cipher and demonstrated that its initial state can be recovered using only six faults. Salam et al. \cite{salametal2021} used bit-flipping and  random fault attacks, under different control models, to recover the  initial state of Grain128AEAD. In a later work, Salam et al. \cite{Salam2023} recovered the initial state and secret key of CLX-128 by applying bit-flipping and random fault attacks with 54 and 134 faults, respectively. Rostami et al. \cite{Rostami2025} applied a random fault attack against Enocoro-80 and Enocoro-128v2. The results show that the attack only require 2 faults to recover the internal state of Enocoro-80 and Enocoro-128v2 with the time complexity of $2^{16}$ and $2^{29}$, respectively. Radheshwar and Roy \cite{Radheshwar2025} applied DFA against ChaosForge stream cipher to recover its internal state and the secret key. Prajasantosa and Salam \cite{Prajasantosa2025} recovered 58 state bits and 15 key bits of TinyJAMBU with bit-flipping and random fault attacks, under different control models.

Recent advancements in neural cryptanalysis have leveraged deep learning (DL) to construct high-performance neural distinguishers for various lightweight block ciphers \cite{gohr2019, houetal2021, zahednejadetal2022}. The deep learning approach has also been used to predict the secret key of lightweight block ciphers, such as S-DES and S-AES \cite{kimetal2023}. In terms of deep learning assisted fault attack, Baksi et al. \cite{baksietal2021} developed multilayer perceptron (MLP) models to identify the fault location for lightweight stream ciphers, Grain128a and Lizard, which achieve better performance than the traditional signature-based method. Cheng et al. \cite{chengetal2023} proposed a deep learning-based fault analysis, using MLP and Convolutional Neural Network (CNN) to recover the key of AES with 1 fault and an average of 1488 ciphertexts. Saha et al. \cite{Saha2023} implemented a fault attack leakage assessment test with deep learning to detect fault-induced leakage in the ciphertext distributions. Mondal et al. \cite{mondal2024} developed a deep neural network for identifying the fault location of Subterranean 2.0 and recovered the embedded secret key with no more than 5 faults.

The existing literature shows that differential fault attacks remain a popular cryptanalytic technique for state recovery or key recovery. In addition, the integration of deep learning in fault attacks offers a new path to enhance the performance of the attack. 
Given the limited research on deep learning–assisted DFA, the feasibility of such fault attacks remains largely underexplored.
To address this, we propose a deep learning–assisted DFA framework for enhancing differential fault attacks and apply it to ACORNv3, MORUSv2, and ATOM.

\subsection{Existing Differential Fault Attacks on ACORN and MORUS} 
Dalai and Roy \cite{dalai2017} implemented differential fault attacks on the first and second version of ACORN. 
By precisely injecting 326 faults and guessing 120 state bits, the full internal state of the cipher at the 39th clock of the encryption phase can be recovered, with an attack complexity of $2^{120}$. Zhang et al. \cite{zhang2017} applied DFA against ACORNv2 under the fault model of no control over the fault location. By analyzing the occurrence of 1s in the differential keystream, the fault locations were identified with a high probability of 97.08\%. They used linearization and guess-and-determine strategies to solve the equations and recover the initial state. The linearization strategy required 31 to 88 faults to recover the initial state with an attack complexity of $c\cdot2^{179.19-1.76n}$, where $n$ is the number of fault injections, and $c$ is the time complexity of solving the equations. For the guess-and-determine method, guessing 67 state bits could recover the initial state of the cipher with 41 fault experiments.

Siddhanti et al. \cite{siddhantietal2017} extended DFA to ACORNv3, optimizing the classical differential framework by incorporating both forward and backward state-update equations and fixing selected bits before SAT solving. In their attack, the fault locations were identified using the signature-based method. Due to the large state size of ACORN, the attack was optimized by guessing 20 state bits. Their work showed that the initial state of ACORNv3 can be recovered using 9 faults and a 1200-bit keystream, with a time complexity of $2^{25.40}$. Zhang et al. \cite{zhangetal2018} also performed DFA on ACORNv3, following the same attack process as in \cite{zhang2017}. To identify the fault location, they introduced the “unique” versus “non-unique” differential sets method. By incorporating the keystream extension and high probability priority strategies, the accuracy of their proposed method was 99.998\%, using 163-bit differential keystream. For equation solving, a guess-and-determine method was used to  recover the full initial state of the cipher. The attack complexity is $c\cdot2^{146.5-3.52n}$, where $n$ is the number of faults such that $26 < n < 43$ faults and $c$ is the time complexity of solving the equations. 

In summary, most existing work on DFAs against ACORN relies on the guess-and-determine method to recover the internal state. 
Furthermore, some of the works focus on retrieving only linear equations, which requires more fault injections to recover the state of the cipher. Additionally, attacks based on the fault model with no control over the fault location rely solely on traditional signature-based methods to identify the fault location. This presents a gap for investigating the use of deep learning to identify the fault locations and improve the attack complexity. 
Furthermore, incorporating higher-degree equations with a threshold-based technique could help optimize the number of required faults.

Wong et al. \cite{Wongetal2020} performed a theoretical analysis of DFA on MORUS, targeting the bit-wise AND operation of the keystream generation function, where a random fault is injected into one of the two involved registers. 
By exploiting these faults, an adversary can reveal the non-zero state bits in the targeted register whenever the differential keystream bits are 1. 
Repeated fault injections allow these non-zero bits to be recovered with high probability, while the remaining bits are assumed to be 0 with high confidence. 
However, the attack only achieved partial state recovery and its reliance on precise control over the fault location is a relatively strong assumption, which motivated us to investigate attaining full state recovery through DL-based models in more relaxed settings. 

\subsection{Our Contributions}
We propose a deep learning-assisted differential fault attack (DL-DFA) framework under the single-bit bit-flipping fault model with no control over the fault location. The contributions of this work are summarized as follows:

\noindent\textbf{MLP Models for Fault Location Identification.} We generate differential keystream data sets for ACORNv3, MORUSv2 and ATOM, and design and train MLP models to identify the fault locations for these ciphers. The models achieve accuracies of 0.999880, 0.999231 and 0.823568 for ACORNv3, MORUSv2 and ATOM, respectively--significantly improving upon the traditional signature-based method. Table~\ref{fault location comparison} summarizes the results of fault location identification in this work.
\begin{table}[t]
\centering
\caption{Summary of fault location identification accuracy across different ciphers}
\label{fault location comparison}
\begin{tabular}{l l c c l}
\toprule
Cipher & Method & Keystream length & Accuracy & Ref. \\
\midrule
\multirow{3}{*}{ACORNv3} 
        & MLP (this work) & 152 bits & 0.999880 & Sec.~\ref{imp dfa acorn} \\
        & Set-based      & 149 bits & 0.999774 & \cite{zhangetal2018} \\
        & Signature-based & 152 bits & 0.999747 & Sec.~\ref{imp dfa acorn} \\
\midrule
\multirow{2}{*}{MORUSv2} 
        & MLP (this work) & 384 bits & 0.999231 & Sec.~\ref{imp dfa morus} \\
        & Signature-based & 384 bits & 0.952606 & Sec.~\ref{imp dfa morus} \\
\midrule
\multirow{2}{*}{ATOM}
        & MLP (this work) & 56 bits  & 0.823568 & Sec.~\ref{imp dfa atom} \\
        & Signature-based & 56 bits  & 0.588976 & Sec.~\ref{imp dfa atom} \\
\bottomrule
\end{tabular}
\end{table}

\noindent\textbf{Threshold-based Attack Framework.} We introduce a threshold-based method to minimize the number of fault injections. Meanwhile, we include all equations into the equation system for state recovery, regardless of the degree. We achieve a full state recovery on ACORNv3 and MORUSv2, improving the existing results in terms of attack complexity and number of recovered bits, respectively. We also implemented the first DFA against ATOM; the results show that ATOM possesses a higher security margin compared to ACORNv3 and MORUSv2. Table~\ref{dfa comparison} summarizes our results of DFA. 
\begin{table}[t]
\centering
\caption{Summary of differential fault attacks on ACORNv3, MORUSv2, and ATOM}
\label{dfa comparison}
\resizebox{\columnwidth}{!}{
\begin{tabular}{l l c l l l}
\toprule
Cipher & Fault type & \# Faults & \# Recovered bits & Complexity & Ref. \\
\midrule
\multirow{3}{*}{ACORNv3} & Bit-flipping & 9 & Full initial state & $2^{25.40}$ & \cite{siddhantietal2017} \\
 & Bit-flipping & 27--42 & Full initial state & $c \cdot 2^{146.5-3.52n}$ & \cite{zhangetal2018} \\
 & Bit-flipping & 21--34 & Full initial state & Negligible & Sec.~\ref{imp dfa acorn} \\
\midrule
\multirow{2}{*}{MORUSv2}  & Random & -- & 256 state bits & -- & \cite{Wongetal2020} \\
 & Bit-flipping & 213--248 & Full initial state & $2^6$ (At most)& Sec.~\ref{imp dfa morus} \\
\midrule
ATOM  & Bit-flipping & 46 & Majority of NFSR state & $\ge 2^{60}$ & Sec.~\ref{imp dfa atom} \\
\bottomrule
\end{tabular}}
\end{table}

\noindent\textbf{Organization of the Paper.} We provide a brief background on fault location identification and an overview of ACORNv3, MORUSv2 and ATOM in Section~\ref{background}. In Section~\ref{attack method}, we discuss the attack procedure, including the implementation of the MLP model and the proposed threshold-based method. The implementation of the attacks and the results for the three ciphers are presented in Section~\ref{imp and result}. Finally, we conclude the paper in Section~\ref{conclusion}. 

\section{Preliminaries and Background}
\label{background}
\subsection{Fault Location Identification}
Our proposed deep learning-assisted fault attack framework works under the fault model of no control over the fault location. In this model, fault location identification is a critical process, as it can affect the secret recovery process. Consider a toy cipher with 4 bits state, $s_0\|s_1\|s_2\|s_3=0\|0\|0\|1$, where the first keystream bit is defined as:
\begin{equation}
\label{fault-free}
    z_0=s_0\oplus s_1 \oplus s_1s_2 \oplus s_2s_3.
\end{equation}
During the attack, let us assume a fault is injected at the location $s_2$, resulting in the value of $s_2$ being complemented. The faulty keystream equation can be rewritten as:
\begin{equation}
\label{faulty}
    z_0^\prime=s_0\oplus s_1 \oplus s_1s_2^\prime \oplus s_2^{\prime}s_3,
\end{equation}
where $s_2^\prime=s_2\oplus1$. By XOR-ing Equation~\ref{fault-free} and Equation~\ref{faulty}, we can obtain the differential:
\begin{equation}
\label{difeq}
\begin{split}
    z_0\oplus z_0^\prime &=(s_0\oplus s_1 \oplus s_1s_2 \oplus s_2s_3) \oplus (s_0\oplus s_1 \oplus s_1s_2^\prime \oplus s_2^{\prime}s_3) \\
    &= s_1s_2 \oplus s_2s_3 \oplus s_1s_2^\prime \oplus s_2^{\prime}s_3 \\
    &= s_1(s_2\oplus s_2^{\prime}) \oplus s_3(s_2\oplus s_2^{\prime}) \\
    &= s_1 \oplus s_3.
\end{split}
\end{equation}
Substituting the values into Equation~\eqref{difeq}, we can find that the expression holds true. Note that in the no-control model, the adversary has no knowledge about the fault location. Suppose the fault is injected at $s_2$, but the adversary misidentifies the fault location as $s_3$, resulting in the following differential equation instead:
\begin{equation}
\begin{split}
    z_0\oplus z_0^\prime &=(s_0\oplus s_1 \oplus s_1s_2 \oplus s_2s_3) \oplus (s_0\oplus s_1 \oplus s_1s_2 \oplus s_2s_3^{\prime}) \\
    &=s_2s_3 \oplus s_2s_3^\prime \\
    &=s_2(s_3 \oplus s_3^{\prime}) \\
    &=s_2.
\end{split}
\end{equation}
Substituting the values into the equation, it shows that the wrong value is revealed for $s_2$ due to the misidentification. Thus, it is important to identify the correct fault location, under the no control model, to ensure that the valid differential can be derived. There are several methods to identify the fault location, with the signature-based method being one of the most popular. One can refer to \cite{maitraetal2015, maitraetal2017, siddhantietal2017} for the details of signature-based method.

\subsection{Specification of Ciphers} 
ACORNv3 is one of the algorithms in the final portfolio of the CAESAR competition for the use case of lightweight applications, while MORUSv2 is one of the additional finalists for high-performance application in the same competition. ATOM, on the other hand, is a stream cipher proposed by Banik et al. \cite{atom} which utilizes a double key filter. In our fault attacks, we mainly focus on the encryption phase of the ciphers, and thus only the encryption process of each cipher is described in the following sections. One can refer to \cite{acorn, morus, atom} for the detailed specification of each cipher.  
\subsubsection{ACORNv3}
Let $S=\{s_0,s_1,\cdots,s_{292}\}$ denote the initial state of ACORNv3. In the encryption phase, the cipher processes the plaintext $P=\{p_0,p_1,\cdots,p_{mlen-1}\}$ and outputs the ciphertext $C=\{c_0, c_1, \cdots,c_{mlen-1}\}$. The cipher utilizes the output function $ks(S)$ to generate a keystream bit $z_i$, where $i=0,1,\cdots,mlen-1$. The output function is defined as:
\begin{equation}
\label{ks}
\begin{split}
        ks(S) &= s_{12}\oplus s_{154}\oplus s_{61}s_{235} \oplus s_{93}s_{235} \oplus s_{61}s_{93} \oplus s_{111}s_{230} \oplus s_{66}\overline{s_{230}},
\end{split}
\end{equation}
where $\overline{s_{230}}$ denotes the complement of $s_{230}$. Additionally, a feedback function $fb(S)$ is used to compute a feedback bit $y$, where the function is defined as
\begin{equation}
\label{fb}
    fb(S)=s_0\oplus\overline{s_{107}}\oplus s_{23}s_{244} \oplus s_{160}s_{244} \oplus s_{23}s_{160} \oplus s_{196},
\end{equation}
where $\overline{s_{107}}$ denotes the complement of the state bit $s_{107}$. The encryption process of ACORNv3 is described in Algorithm~\ref{encACORN}.
\begin{algorithm}[t]
\caption{Encryption phase of ACORNv3}
\label{encACORN}
\begin{algorithmic}
    \STATE Initialize list $Z$
    \FOR{$i=0$ \textbf{to} $mlen-1$}
    \STATE $s_{289}=s_{289}\oplus s_{235}\oplus s_{230}$
    \STATE $s_{230}=s_{230}\oplus s_{196}\oplus s_{193}$
    \STATE $s_{193}=s_{193}\oplus s_{160}\oplus s_{154}$
    \STATE $s_{154}=s_{154}\oplus s_{111}\oplus s_{107}$
    \STATE $s_{107}=s_{107}\oplus s_{66}\oplus s_{61}$
    \STATE $s_{61}=s_{61}\oplus s_{23}\oplus s_0$
    \STATE $z_i = ks(S)$
    \STATE Append $z_i$ to $Z$
    \STATE $y = fb(S)$
    \FOR{$j=0$ \textbf{to} $291$}
    \STATE $s_j = s_{j+1}$
    \ENDFOR
    \STATE $s_{292} = y \oplus p_i$
    \ENDFOR
    \RETURN $Z=\{z_0,z_1,\cdots,z_{mlen-1}\}$
\end{algorithmic}
\end{algorithm}
The state bits $s_{61}$, $s_{107}$, $s_{154}$, $s_{193}$, $s_{230}$ and $s_{289}$ first undergo an intermediate update. 
Next, the keystream and feedback bits are computed based on this intermediate state. 
Finally, the cipher updates the internal state by shifting, the last state bit $s_{292}$ is updated by XOR-ing the feedback bit $y$ with the plaintext bit $p_i$. 
\subsubsection{MORUSv2}
We apply the attack on MORUS-640-128, which is one of the three variants of MORUSv2. It takes a 128-bit secret key, with an internal state of 640 bits. The state is constructed of 5 registers, denoted as $S=S_0\|S_1\|S_2\|S_3\|S_4$, where each register holds 128-bit data. In each step of the encryption phase, the cipher processes a 128-bit plaintext block. Let the plaintext $P$ consist of $y$ blocks, where the plaintext block is denoted as $P_i$, $i \in \{0,1,\cdots,y-1\}$. If the last plaintext block $P_{y-1}$ is not a full block, a 0 valued string is padded to that block. The keystream block $Z_i$ is computed by $S_0 \oplus (S_1 \ll 96) \oplus S_2S_3$, where $S_1$ is rotated to the left by 96 bits, as defined in Algorithm~\ref{encMORUS}.
\begin{algorithm}[!htb]
\caption{Encryption phase of MORUSv2}
\label{encMORUS}
\begin{algorithmic}
    \STATE Initialize list $Z$
    \FOR{$i=0$ \textbf{to} $y-1$}
    \STATE $Z_i=S_0 \oplus (S_1 \ll 96) \oplus S_2S_3$
    \STATE Append $Z_i$ to $Z$
    \STATE Update the state with state update function
    \ENDFOR
    \RETURN $Z=\{Z_0, Z_1, \cdots, Z_{y-1}\}$
\end{algorithmic}
\end{algorithm}
\subsubsection{ATOM}
ATOM consists of a linear feedback shift register (LFSR) and a non-linear feedback shift register (NFSR). Let the initial states of the LFSR and NFSR be $L=\{l_0,l_1,\cdots, l_{68}\}$ and $B=\{b_0,b_1,\cdots,b_{89}\}$, respectively. During encryption, the keystream bit is computed as: 
\begin{equation}
\begin{split}
    O\left(B,L\right) &=b_1\oplus b_5\oplus b_{11}\oplus b_{22}\oplus b_{36}\oplus b_{53}\oplus b_{72}\oplus b_{80} \oplus b_{84} \oplus l_5l_{16} \oplus l_{13}l_{15} \\& \oplus l_{30}l_{42} \oplus l_{22}l_{67} \oplus h\left(l_7,l_{33},l_{38},l_{50},l_{59},l_{62},b_{85},b_{41},b_9\right).
\end{split}
\end{equation}
Note that $h\left(l_7,l_{33},l_{38},l_{50},l_{59},l_{62},b_{85},b_{41},b_9\right)$ is a degree 5 function. Both the LFSR and NFSR possess the corresponding feedback function to update the state. The feedback functions for the LFSR and NFSR are denoted by $F\left(L\right)$ and $G\left(B\right)$, respectively, and are defined as:
\begin{equation}
    F\left(L\right)=l_0\oplus l_5\oplus l_{12}\oplus l_{22}\oplus l_{28}\oplus l_{37}\oplus l_{45}\oplus l_{58},
\end{equation}
\begin{equation}
\begin{split}
    G\left(B\right)&=b_0\oplus b_{24}\oplus b_{49}\oplus b_{79}\oplus b_{84}\oplus b_3b_{59}\oplus b_{10}b_{12} \oplus b_{15}b_{16}\oplus b_{25}b_{53} \\& \oplus b_{35}b_{42} \oplus b_{55}b_{58} \oplus b_{60}b_{74} \oplus b_{20}b_{22}b_{23}\oplus b_{62}b_{68}b_{72}\oplus b_{77}b_{80}b_{81}b_{83}.
\end{split}
\end{equation}
Algorithm~\ref{encATOM} defines the encryption phase of ATOM. Note that the last seven bits in the LFSR are interpreted as a decimal number, $cnt$, which determines the index of the key bit $k$ involved in the computation of the feedback bit for the NFSR. Specifically, the NFSR feedback bit is computed by XOR-ing the output of update function with $l_0$ and two key bits chosen through the $cnt$ and modulo of current clock. The value of $cnt$ is non-deterministic, as it depends on the evolving state of the LFSR. 
\begin{algorithm}[t]
\caption{Encryption phase of ATOM}
\label{encATOM}
\begin{algorithmic}
    \STATE Initialize list $Z$
    \FOR{$i=0$ \textbf{to} $mlen-1$}
    \STATE $z_i= O(B,L)$
    \STATE Append $z_i$ to $Z$
    \STATE $cnt=l_{62}\|l_{63}\|l_{64}\|l_{65}\|l_{66}\|l_{67}\|l_{68}$
    \STATE $fb\_B=G\left(B\right)\oplus l_0\oplus k_{cnt}\oplus k_{511+i\ mod\ 128}$
    \FOR{$j=0$ \textbf{to} 88}
        \STATE $b_j=b_{j+1}$
    \ENDFOR
    \STATE $b_{89}=fb\_B$
    \STATE $fb\_L=F\left(L\right)$
    \FOR{$j=0$ \textbf{to} 67}
        \STATE $l_j=l_{j+1}$
    \ENDFOR
    \STATE $l_{68}=fb\_L$
    \ENDFOR
    \RETURN $Z=\{z_0,z_1,\cdots,z_{mlen-1}\}$
\end{algorithmic}
\end{algorithm}
\section{Attack Description}
\label{attack method}
Our attacks are conducted in a simulation-based environment, where the fault injection is simulated via software rather than on a physical device. The attack assumes a known-plaintext attack (KPA) model, where the adversary has access to specific keystream bits.
Furthermore, the following assumptions are made:
\begin{itemize}
    \itemsep0em 
    \item The adversary can inject a single-bit bit-flipping fault into the initial state
    \item The adversary has no control over the fault location. 
    \item The adversary can reset the cipher and run the cipher with the embedded key and initialization vector (IV) to observe the keystream.
\end{itemize}
The attack consists of two phases, the fault location identification and secret recovery. The details of these two phases are discussed in the following section.

\subsection{Fault Location Identification}
The fault location identification is the critical process in our attack, as the fault is injected at an unknown location. Without knowledge of the fault location, we cannot derive valid differential keystream equations to recover the secret of the cipher. To address this, we trained MLP models to identify the fault location using differential keystream datasets. 

\subsubsection{Dataset Generation}
Since no public datasets are available for fault location identification for these ciphers, we generated the datasets ourselves. The procedure of dataset generation for each cipher is defined in Algorithm~\ref{gendata}.
\begin{algorithm}[htbp]
\caption{Dataset Generation}
\label{gendata}
\begin{algorithmic}
    \FOR{each possible fault location $f$ of the cipher}
        \FOR{$j=1$ \textbf{to} 1536}
        \STATE Generate random key $K$ and IV $V$
        \STATE Run the cipher and obtain fault-free keystream $Z$
        \STATE Reset the cipher to initial state
        \STATE Inject bit-flipping fault at location $f$
        \STATE Run the cipher and obtain faulty keystream $Z^{\prime}$
        \STATE Obtain differential keystream $\Delta Z=Z\oplus Z^{\prime}$
        \IF{$j\le 1024$}
        \STATE Write $K$, $V$, $Z$, $Z^{\prime}$, $\Delta Z$, $f$ to training.csv
        \ELSIF{$j \le 1280$}
        \STATE Write $K$, $V$, $Z$, $Z^{\prime}$, $\Delta Z$, $f$ to testing.csv
        \ELSE
        \STATE Write $K$, $V$, $Z$, $Z^{\prime}$, $\Delta Z$, $f$ to validation.csv
        \ENDIF
        \ENDFOR
    \ENDFOR
\end{algorithmic}
\end{algorithm}
For each possible fault location $f$ of the cipher, 1536 samples are generated, each using a random key, $K$ and IV, $V$. The key-IV pair is used to run the cipher for producing the fault-free keystream $Z$. Subsequently, the cipher is reset and run with the same key-IV pair, but a bit-flipping fault is injected to location $f$ in the initial state to produce the faulty keystream $Z^\prime$. The differential keystream is then computed by XOR-ing $Z$ with $Z^\prime$. Finally, the key-IV pair, fault-free keystream, faulty keystream, differential keystream and the fault location are recorded into the csv files. 

For each cipher, three datasets are generated: training, testing, and validation. To enhance the neural network's ability to capture fault propagation patterns, the train-test ratio is set as 80:20. The size of training dataset is set as $n\cdot2^{10}$, while testing and validation datasets are $n\cdot2^8$, where $n$ is the number of possible fault locations. Hence, each fault location has $2^{10}$ samples for training,  $2^8$ samples for testing. The validation dataset is used to evaluate the performance of the model during the training. The length of the differential keystream is set to the range between 50\% and 60\% of the state size of the cipher, as empirical studies suggest that this range is sufficient to capture the unique fault propagation signatures associated with different fault locations.

\subsubsection{Implementation of Multilayer Perceptron Model}
\label{model imp method}
We employ multilayer perceptron (MLP) models to identify the fault locations.
The models were implemented with the Keras library provided in the TensorFlow framework on the Google Colab platform. The number of neurons in the input layer is the length of the differential keystream, while the number of neurons in the output layer is the size of the internal state of the cipher, i.e., the number of possible fault locations. 

The model is compiled with the Adam optimizer and the sparse categorical cross-entropy loss function. As this is a multi-class classification problem, accuracy is used as the primary evaluation metric. The model is trained for a maximum of 50 epochs, with an early-stopping criterion based on validation accuracy. If no improvement in validation accuracy is observed for seven consecutive epochs, the training process is terminated, and the weights associated with the best validation accuracy are restored. 
Various hyperparameters, including the number of hidden layers, number of neurons per layer, dropout rate, and initial learning rate, are chosen through a systematic search over selected configurations.

\subsection{Secret Recovery}
\begin{figure}[htbp]
    \centering
    \includegraphics[width=0.95\linewidth]{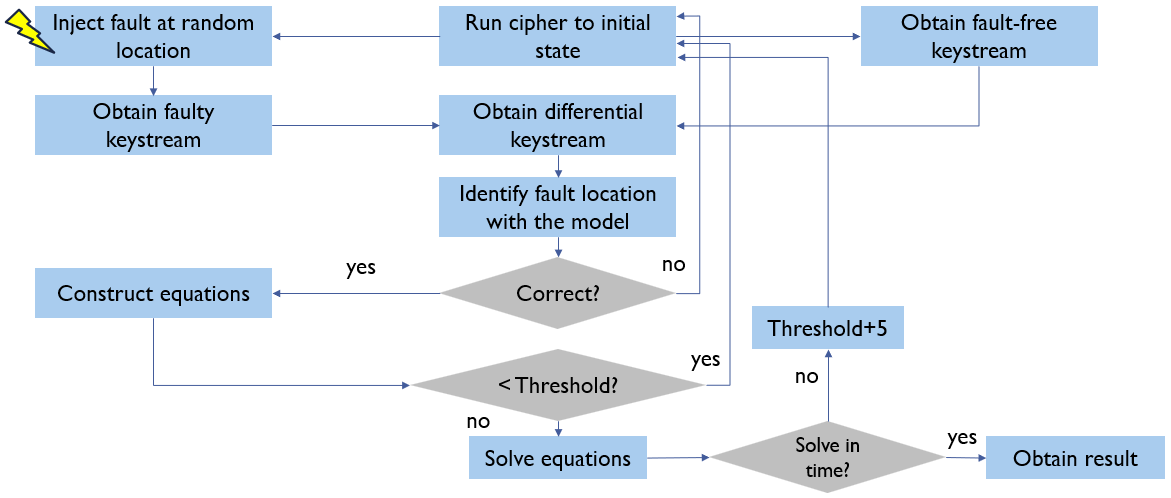}
    \caption{Threshold-based secret recovery process.}
    \label{recoverprocess}
\end{figure}

Once the MLP models are trained, we proceed to apply the differential fault attack to the targeted ciphers. Figure~\ref{recoverprocess} illustrates the process of the state recovery method. The attack begins by obtaining the fault-free keystream of the cipher. 
The cipher is then reset to its initial state, and a single bit-flipping fault is injected at an unknown location. 
We then run the cipher with the faulty initial state to obtain the faulty keystream. 
The differential keystream is then fed into the trained MLP model to identify the fault location. If the fault location is correctly identified, we proceed to the equation construction stage. Conversely, if the fault location is misidentified, the cipher is reset and the process is repeated. 

With the correct fault location, we generate the corresponding differential keystream equations. 
A threshold is introduced to determine the minimum number of differential equations required for the state recovery while minimizing the number of fault injections. If the number of differential equations is below the defined threshold, the fault injection process is repeated to gather more information. 
%
Once the number of equations reaches or exceeds the threshold, we attempt to solve the system using the Gr\"obner basis method provided by the SageMath library. 
In our experiments, we set a one-minute time limit for the solver. 
If the Gr\"obner basis returns the result within this time frame, we then verify the recovered secret bits. Otherwise, the current computation is terminated, the threshold is increased by 5 and the process 
is repeated. Increasing the threshold adds more differential equations to the system, providing more information to the solver and increasing the likelihood of a successful recovery, but also increases the number of faults.
%
\section{Experimental Evaluation and Analysis of the Attacks}
\label{imp and result}
The implementation\footnote{Source codes and datasets are available via GitHub: \url{https://github.com/kokping0605/Deep_Learning_Assisted_Fault_Attack_Framework.git}} 
of the attack framework and the results for each cipher are discussed in this section. Note that the attacks on ACORN and MORUS follow the attack assumption defined in Section~\ref{attack method}. For the attack on ATOM, there is a minor change in attack assumptions, which we will restate in the corresponding section. Due to the varying structure of these ciphers, the model architectures and state recovery processes utilize different settings; these are discussed in detail in the following sections.

\subsection{Differential Fault Attack on ACORNv3}
\label{imp dfa acorn}
For the attack on ACORN, the adversary is assumed to have knowledge of a 152-bit keystream. 
In this attack, the adversary obtains the differential keystream and feeds it into the trained MLP model to identify the fault location and generate the corresponding differential equations with respect to the initial state of the cipher. Once a sufficient number of equations is collected, they are solved to recover the initial state of ACORNv3. 

\subsubsection{Implementation of MLP Model for Fault Location Identification}
Before conducting the attack, three datasets, namely training, testing and validation datasets are generated using Algorithm~\ref{gendata} to train the MLP model for fault location identification in ACORNv3. Training and hyperparameter tuning follow the method defined in Section~\ref{model imp method}. The architecture of the neural network is described as follows:
\begin{itemize}
\itemsep0em
    \item \textbf{Input layer:} 
    The number of neurons in the input layer is 152.
    \item \textbf{Hidden layer:} Five hidden layers, with 152, 512, 512, 512, and 512 neurons, respectively, are used. Each is a dense layer applying a fully connected transformation with $L_2$ weight decay $\left(\lambda=1\times{10}^{-6}\right)$ to prevent overfitting, followed immediately by batch normalization to normalize intermediate activations and mitigate internal covariate shift. The ELU activation function is used to combine fast convergence with robustness to vanishing gradients and a dropout rate of 0.3 is set after each activation to prevent co-adaptation of neurons and reduce overfitting. 
    \item \textbf{Output layer:} The number of neurons in this layer is 293, and they are activated with the softmax activation function.
\end{itemize}
\subsubsection{Implementation of Differential Fault Attack on ACORN}
The detailed attack process of ACORN is defined in Algorithm~\ref{dfa acorn}. 
Initially, two lists $\mathcal{E}$ and $\mathcal{F}$, are defined to store the differential equations and identified fault locations, respectively. The threshold $\mathcal{T}$ determines the minimum number of linear equations required in the system. Through our observation of different threshold settings, we found that $\mathcal{T}=150$ provides the best results for the state recovery attack. 
Prior to the attack, the adversary obtains the 152-bit fault-free keystream, $Z=\{z_0,z_1,\cdots,z_{151}\}$ and generates fault-free Algebraic Normal Form (ANF) with respect to the initial state. 
\begin{algorithm}[!t]
\caption{Differential Fault Attack on ACORNv3}
\label{dfa acorn}
\begin{algorithmic}
    \STATE Initialize list $\mathcal{E}, \mathcal{F}$
    \STATE $solve \gets$\FALSE, $\mathcal{T}\gets150$, $num\_deg1\gets0$
    \STATE $Z=\{z_0, z_1, \cdots,z_{151}\}\gets encACORN(S, P)$
    \STATE Generate fault-free ANF, $NKS=\{nks_0,\cdots,nks_{151}\}$
    \WHILE{$solve$ \textbf{is} \FALSE}
        \WHILE{$num\_deg1<\mathcal{T}$}
            \STATE Reset cipher to initial state $S^\prime$
            \STATE Inject fault at location $f$, where $s_f=s_f\oplus1$
            \STATE $Z^\prime ={{z}_0^\prime, z_1^\prime, \cdots,z_{151}^\prime}\gets encACORN(S\prime,P)$
            \FOR{$i=0$ \textbf{to} 151}
                \STATE $\Delta z_i^f = z_i \oplus z_i^\prime$
            \ENDFOR
            \STATE $f^\prime\gets modelACORN(\Delta Z^f=\{\Delta z_0^f,\cdots,\Delta z_{151}^f\})$ 
            \IF{$f$ \textbf{not in} $\mathcal{F}$ \AND $f=f^\prime$}
            \STATE Append $f$ to $\mathcal{F}$
            \ELSE
            \STATE \textbf{continue}
            \ENDIF
            \STATE Generate faulty ANF, $FKS=\{fks_0,\cdots,fks_{151}\}$
            \FOR{$i=0$ \textbf{to} 151}
                \STATE $dks=nks_i\oplus fks_i\oplus\mathrm{\Delta}z_i^f$
                \IF{$dks.deg=1$ \AND $dks$ \textbf{not in} $\mathcal{T}$}
                \STATE Append $dks$ to $\mathcal{E}$
                \STATE $num\_deg1 = num\_deg1 + 1$
                \ELSIF{$dks.deg=2$ \AND $dks$ \textbf{not in} $\mathcal{T}$}
                \STATE Append $dks$ to $\mathcal{E}$
                \ENDIF
            \ENDFOR
        \ENDWHILE
        \STATE Solve $\mathcal{E}$ with Gr\"obner basis
        \IF{Gr\"obner basis can solve in time}
            \STATE $solve \gets$\TRUE
        \ELSE
            \STATE $\mathcal{T}=\mathcal{T}+5$
        \ENDIF
    \ENDWHILE
\end{algorithmic}
\end{algorithm}

In the online phase, a bit-flipping fault is injected at a random location $f$ in the initial state, such that $s_f$ is complemented. The adversary utilizes the resulting faulty initial state to obtain the faulty keystream, $Z^\prime =\{{z}_0^\prime,\ z_1^\prime,\ \cdots,\ z_{151}^\prime\}$, and computes the differential keystream, $\Delta Z^f=\{\Delta z_0^f,\Delta z_1^f,\cdots,\Delta z_{151}^f\}$ by $\Delta z_i^f=z_i\oplus z_i^\prime$, where $i=0,1,\cdots,151$. The differential $\Delta Z^f$ is fed to the MLP model to identify the fault location. If the fault location $f$ is correctly identified and is not already present in $\mathcal{F}$, it is added to the list. If the location has already been identified, the adversary re-injects a fault at a different random location. This ensures that only unique fault locations contribute to the equation system.

Once a unique location is identified, the faulty ANFs with respect to the initial state of the cipher, $FKS=\{fks_0,fks_1,\cdots,fks_{151}\}$ are generated. The adversary then derives the differential equations, $dks = nks_i\oplus fks_i\oplus\Delta z_i^f$, where $i=\{0,1,\cdots,151\}$. Each time a unique linear differential equation is obtained, the counter $num\_deg1$ is increased by 1 and the equation is added into the equations list $\mathcal{E}$. The obtained unique quadratic equation is also added to the same list $\mathcal{E}$. This process is repeated iteratively until $num\_deg1$ is greater than or equal to the threshold, $\mathcal{T}$. 
Subsequently, the system of differential equations in $\mathcal{E}$ are solved using Gr\"obner basis to recover the initial state $S$. 
\subsubsection{Results and Discussions of the Attack on ACORNv3}
The performance of the trained model is shown in Table~\ref{acorn model performance}. 
Utilizing the proposed MLP architecture, the model achieves a high accuracy of 99.9880\%, showing the model has successfully learned the fault propagation patterns to identify the fault locations with 152-bit differential keystream. The precision score of 99.9880\% shows that the model is robust and effective to identify the actual fault locations with a negligible false positive rate. In terms of recall of 99.9880\%, it shows that the model is able to capture nearly all fault locations.
The F1-score of 99.9880\% demonstrated balanced precision-recall across all target locations, validating the architectural efficacy and hyperparameter configuration.
\begin{table}[!htbp]
\centering
\caption{Performance of trained model for identifying fault locations on ACORNv3}
\label{acorn model performance}
\begin{tabular}{cccc}
\hline
Accuracy & Precision & Recall & F1-score \\
\hline
0.999880 & 0.999880 & 0.999880 & 0.999880 \\
\hline
\end{tabular}
\end{table}

Furthermore, we benchmarked the performance of our model against the set-based fault location identification method by Zhang et al. \cite{zhangetal2018}. Additionally, we implemented a traditional signature-based method using our generated dataset, where the training data were used to generate signatures and the testing data were used for evaluation. Because the differential keystream used by our model is longer than the 99-bit keystream used by Zhang et al. \cite{zhangetal2018}, we performed the comparison based on the keystream extension method described in their work. 
For the 149-bit keystream extension case, our model achieves an accuracy that is 0.0106\% higher than that of the set-based method. Compared to the traditional signature-based method, which achieved an accuracy of 99.9747\%, our proposed MLP model provides a 0.0133\% improvement. These results confirm that the MLP-based approach offers superior and more robust fault location identification.

A total of 400 experiments were conducted to analyze the performance of the proposed method in recovering the initial state of ACORNv3. Excluding eight experiments exhibiting misidentified fault locations, the remaining trials demonstrated that the 293-bit initial state of ACORNv3 can be fully recovered. The attack was successful with a high probability of 0.98. While experiments were halted upon misidentification, the model still identified the majority of fault locations correctly in those cases. This suggests that even with minor identification errors, a substantial portion of the state could likely be recovered, as inconsistent equations can be filtered out through satisfiability checking.

\begin{figure}[!tb]
    \centering
    \includegraphics[width=0.7\linewidth]{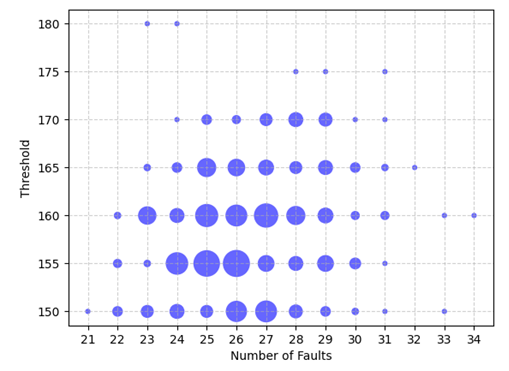}
    \caption{Correlation between fault injection counts and equation thresholds for ACORNv3}
    \label{t and f}
\end{figure}

Figure~\ref{t and f} illustrates the relationship between the thresholds and the number of faults required to recover the initial state of ACORNv3. 
Experimental results show that the number of faults ranges between 21 and 34, with an average of 27 injections required for full recovery. In Figure~\ref{t and f}, the larger the circle, the more experiments utilized the same number of faults and threshold. The figure shows that the same number of faults may result in a varied range of thresholds, as different fault locations can generate different number of linear equations. 

\begin{figure}[!tb]
    \centering
    \includegraphics[width=\linewidth]{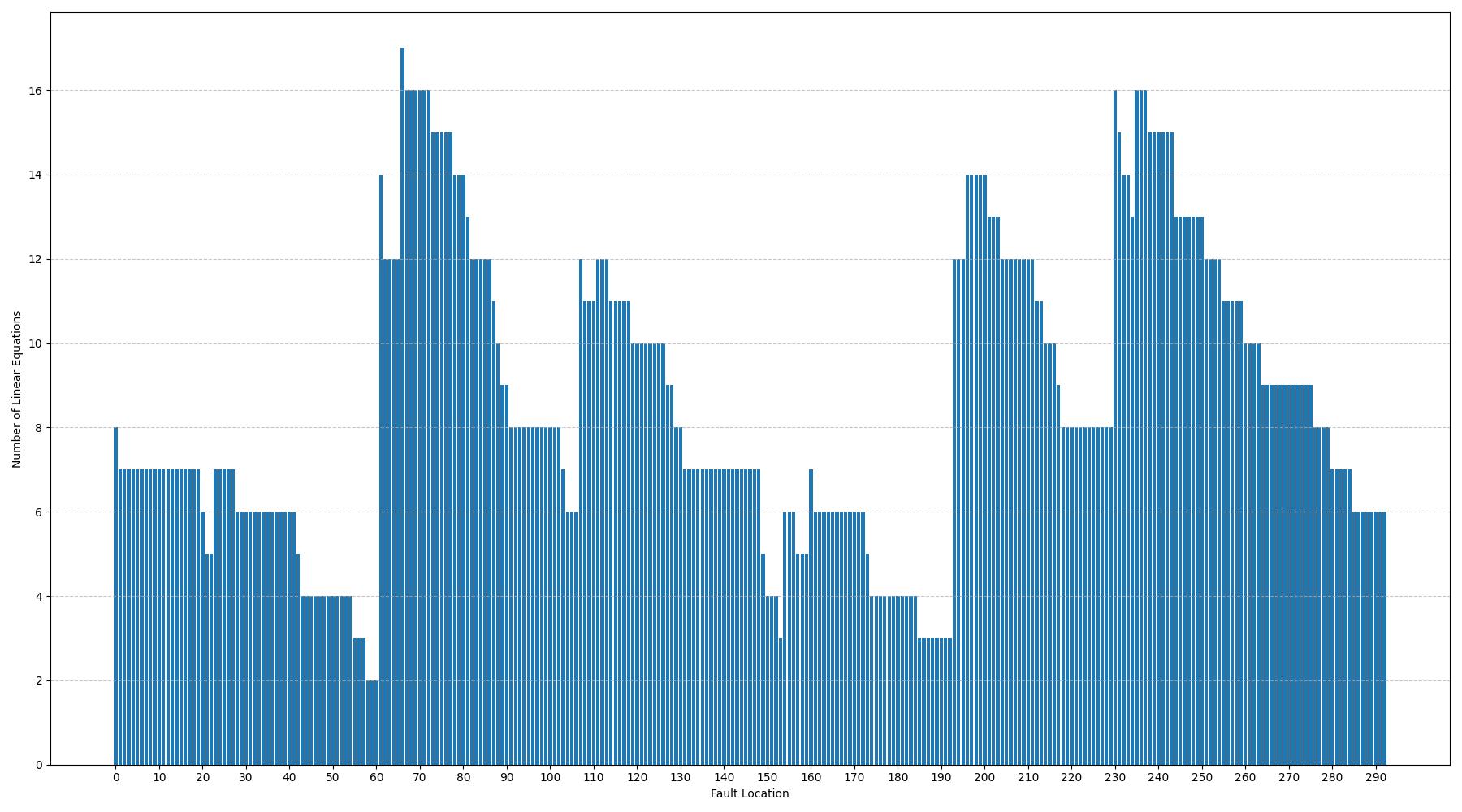}
    \caption{Number of linear equations produced by each fault location in ACORNv3.}
    \label{fault and linear acorn}
\end{figure}

Figure~\ref{fault and linear acorn} shows the distribution of linear differential equations produced by each fault location in ACORNv3. It is observed that some locations, such as 66 to 77, 230, 231 and 235 to 243 generate 15 or above linear equations, which are the ideal fault locations. If the fault is injected at these locations, the number of faults required in the attack can be reduced. However, the required number of fault injections also depends on the algebraic complexity of the resulting equations. If the system is too complex for the Gr\"obner basis sovler to solve within the set time limit, additional fault injections are required. 

For the study conducted by Siddhanti et al. \cite{siddhantietal2017}, although only a small number (9) of faults are required to recover the state, the attacker must know the values of 20 state bits, which increases the complexity of the attack by $2^{20}$. In the work by Zhang et al. \cite{zhangetal2018}, the authors focus on recovering the initial state of ACORNv3 by solving the system of linear equations. Their approach requires $26 < n < 43$ faults to recover the initial state using a guess-and-determine method, as the number of linear equations obtained is fewer than the number of state variables. Because their attack requires guessing certain state bits, the attack complexity is $c\cdot2^{146.5-3.52n}$, where $c$ is the time complexity of equation solving and $n$ is the number of faults. 

In contrast, we incorporate both linear and quadratic equations for the state recovery process, and achieve a full state recovery without necessitating a guess-and-determine approach. Our proposed method successfully recovers the full 293-bit state of ACORNv3 with 21 to 34 faults. 
This represents a 19.04\% reduction in the maximum number of faults required compared to the results of Zhang et al. \cite{zhangetal2018}.
The time complexity of our attack is negligible and significantly lesser than the two existing works. 

\subsection{Differential Fault Attack on MORUSv2}
\label{imp dfa morus}
For the attack on MORUS, the adversary is assumed to have the knowledge of three plaintext blocks $P=\{P_0,P_1,P_2\}$, i.e., a total of a 384-bit plaintext. 
During the attack, the adversary obtains the differential keystream and feeds it into the trained model to identify the fault location and generate the corresponding differential equations with respect to the initial state. The same process is repeated until sufficient number of differential equations are collected. The initial state of MORUS is recovered by solving the equations. In the following sections, the initial state is denoted as $S\ =\{s_0,s_1,\cdots,s_{639}\}$. 	

\subsubsection{Implementation of MLP Model for Fault Location Identification}
Using the datasets generated with Algorithm~\ref{gendata}, an MLP model for fault location identification for MORUS is implemented. We train the model and tune the hyperparameters with the method defined in Section~\ref{model imp method}. The model architecture is described as follow:
\begin{itemize}
\itemsep0em
    \item \textbf{Input layer:} This input layer consists of 384 neurons.
    \item \textbf{Hidden layer:} A total of three hidden layers are used, each containing 512 neurons. Each is a dense layer followed by batch normalization to mitigate internal covariate shift and increase training stability. For the first two hidden layers, the ReLU activation function and the dropout rate of 0.2 are applied after the batch normalization to reduce overfitting. The final hidden layer utilizes ReLU activation after batch normalization without dropout.
    \item \textbf{Output layer:} This layer consists of 640 neurons (the number of possible fault locations in MORUS). We utilize softmax as the activation function in this layer. 
\end{itemize}

\subsubsection{Implementation of Differential Fault Attack on MORUS}
The process of differential fault attack on MORUS is described in Algorithm~\ref{dfa morus}. 
The initial settings of the attack are identical to the attack done on ACORN, except for the setting of the threshold $\mathcal{T}$. The threshold $\mathcal{T}$ is obtained through experimental observation, as in the case of ACORNv3, and set to 896 for MORUS. To apply the attack to MORUS, we include all equations into the equation system, regardless the degree of equations. 
Prior to the attack, the adversary first runs the cipher with fault-free initial state $S=\{s_0,s_1,\cdots,s_{639}\}$ and obtains the 384-bit fault-free keystream, $Z=\{z_0,z_1,\cdots,z_{383}\}$. Due the constraint of SageMath, we can only generate the ANF for the first 256 iterations of the encryption phase; therefore we construct 256 fault-free keystream equations $NKS=\{nks_0,nks_1,\cdots,nks_{255}\}$$NKS=\{nks_0,nks_1,\cdots,nks_{255}\}$.

In the online phase of the attack, it follows the same procedure as in the attack on ACORNv3, where the adversary reset the cipher to the initial state, $S^\prime=\{s_0,s_1,\cdots,s_{639}\}$ and injects a bit-flipping fault at a random location $f$, such that $s_f$ is complemented. The faulty initial state is then used to run the cipher and output the faulty keystream, $Z^\prime\ ={{z}_0^\prime,\ z_1^\prime,\ \cdots,\ z_{383}^\prime}$. The resulting differential keystream, $\Delta Z^f={\Delta z_0^f,\Delta z_1^f,\cdots,\Delta z_{383}^f}$, is fed into the MLP model to identify the fault location. Once the fault location is identified, the ANFs of the first 256-bit faulty keystream with respect to the initial state of the cipher are generated. Each retrieved differential equation is added to the equation system $\mathcal{E}$,  regardless of its algebraic degree. The process is repeated until the number of equations in the equation system reaches or exceeds the threshold of 896. 
\begin{algorithm}[!t]
\caption{Differential Fault Attack on MORUSv2}
\label{dfa morus}
\begin{algorithmic}
    \STATE Initialize list $\mathcal{E}, \mathcal{F}$
    \STATE $solve \gets$\FALSE, $\mathcal{T}\gets896$, $num\_eq\gets0$
    \STATE $Z=\{z_0, z_1, \cdots,z_{383}\}\gets encMORUS(S, P)$
    \STATE Generate fault-free ANF, $NKS=\{nks_0,\cdots,nks_{255}\}$
    \FOR{$i=0$ \textbf{to} 255}
        \STATE Append $nks_i\oplus z_i$ to $\mathcal{E}$
    \ENDFOR
    \WHILE{$solve$ \textbf{is} \FALSE}
        \WHILE{$num\_eq<\mathcal{T}$}
            \STATE Reset cipher to initial state $S^\prime$
            \STATE Inject fault at location $f$, where $s_f=s_f\oplus1$
            \STATE $Z^\prime =\{{z}_0^\prime, z_1^\prime, \cdots,z_{383}^\prime\}\gets encMORUS(S^\prime,P)$
            \FOR{$i=0$ \textbf{to} 383}
                \STATE $\Delta z_i^f = z_i \oplus z_i^\prime$
            \ENDFOR
            \STATE $f^\prime\gets modelMORUS(\Delta Z^f=\{\Delta z_0^f,\cdots,\Delta z_{383}^f\})$ 
            \IF{$f$ \textbf{not in} $\mathcal{F}$ \AND $f=f^\prime$}
            \STATE Append $f$ to $\mathcal{F}$
            \ELSE
            \STATE \textbf{continue}
            \ENDIF
            \STATE Generate faulty ANF, $FKS=\{fks_0,\cdots,fks_{255}\}$
            \FOR{$i=0$ \textbf{to} 255}
                \STATE $dks=nks_i\oplus fks_i\oplus\mathrm{\Delta}z_i^f$
                \IF{$dks.deg\ge1$ \AND $dks$ \textbf{not in} $\mathcal{T}$}
                \STATE Append $dks$ to $\mathcal{E}$
                \STATE $num\_eq = num\_eq + 1$
                \ENDIF
            \ENDFOR
        \ENDWHILE
        \STATE Solve $\mathcal{E}$ with Gr\"obner basis
        \IF{Gr\"obner basis can solve in time}
            \STATE $solve \gets$\TRUE
        \ELSE
            \STATE $\mathcal{T}=\mathcal{T}+5$
        \ENDIF
    \ENDWHILE
\end{algorithmic}
\end{algorithm}

Once a sufficient number of differential equations has been collected and stored in $\mathcal{E}$, a Gr\"obner basis is applied to solve the system and recover the initial state. This procedure follows a similar approach to that used for ACORN. Note that in this attack, we also include the first 256 bits of the original keystream equations into the equation system $\mathcal{E}$ to make it more constrained, due to the larger state size of MORUS. 

\subsubsection{Results and Discussions of the Attack on MORUSv2}
The performance of the trained MLP model in fault location identification is shown in Table~\ref{morus model performance}. The results in Table~\ref{morus model performance} show that the trained model achieves a high accuracy of 99.9231\%, i.e., the model is able to capture most of the pattern of fault propagation. 
With a precision score of 99.9235\%, the model demonstrates a robust ability to correctly classify fault locations while maintaining a very low probability of false positive identifications.
The recall score of 99.9231\% is slightly lower than the precision score by 0.0004\%, indicating that the probability of misidentification is slightly higher than the probability of false positive identifications. However, the score of 99.9231\% is still considered as a good score, as the model can identify most of the actual fault locations among all actual positives cases. The F1 score of 99.9231\% shows a balanced trade-off between the precision and recall, indicating that the trained model has high reliability in identifying the fault locations.
\begin{table}[!t]
\centering
\caption{Performance of trained model for identifying fault locations on MORUSv2}
\label{morus model performance}
\begin{tabular}{cccc}
\hline
Accuracy & Precision & Recall & F1-score \\
\hline
0.999231 & 0.999235 & 0.999231 & 0.999231 \\
\hline
\end{tabular}
\end{table}

To provide a comprehensive evaluation, we also implemented a traditional signature-based method to benchmark the performance of the trained MLP model. 
The signature-based method utilized the same training and testing datasets as the MLP model for signature generation and evaluation, respectively.
The accuracy of fault location identification for the MLP model is 99.9231\%, while the signature-based method is 95.2606\%. The result shows that the proposed MLP model outperforms the traditional signature-based method in fault location identification of MORUS. The state update function of MORUS utilizes AND, XOR, and rotation operations, which complicates the fault propagation through the state. The MLP model demonstrates a superior ability to learn these complex, non-linear propagation patterns.
In contrast, the signature-based method relies on static, predefined signatures, and may struggle with the diffusion properties of the cipher. In certain locations, fault propagation becomes computationally indistinguishable, leading to identification errors that the neural network is better equipped to resolve.

We conducted a total of 400 experiments to analyze the performance of the proposed state recovery attack on MORUSv2. 
The results show that 76 out of 400 experiments encountered a misidentification of the fault location.
While the trained model maintains a high accuracy in fault location identification, the large state size of MORUS requires a larger number of faults (213 to 248 faults) to recover the initial state. The increase in number of faults has degraded the performance of model, as the higher number of fault injections has also increased the probability of misidentification cases. 
Note that in these 76 misidentification cases, the model still correctly identified the majority of fault locations within each experiment. However, the experimental setup was designed to terminate prior to the equation-solving phase upon the first occurrence of a misidentification. This conservative approach ensures that the reported success rates reflect a "zero-tolerance" policy for identification errors, even though many of these "failed" experiments contained only a single erroneous location out of more than 200 injections.
Because the number of correct identifications outweighs the erroneous ones, the majority of the derived equations remain valid. Consequently, it is expected that solving the system of equations with misidentification cases can recover the majority of the state bits.

\begin{figure}[!htbp]
    \centering
    \includegraphics[width=0.9\linewidth]{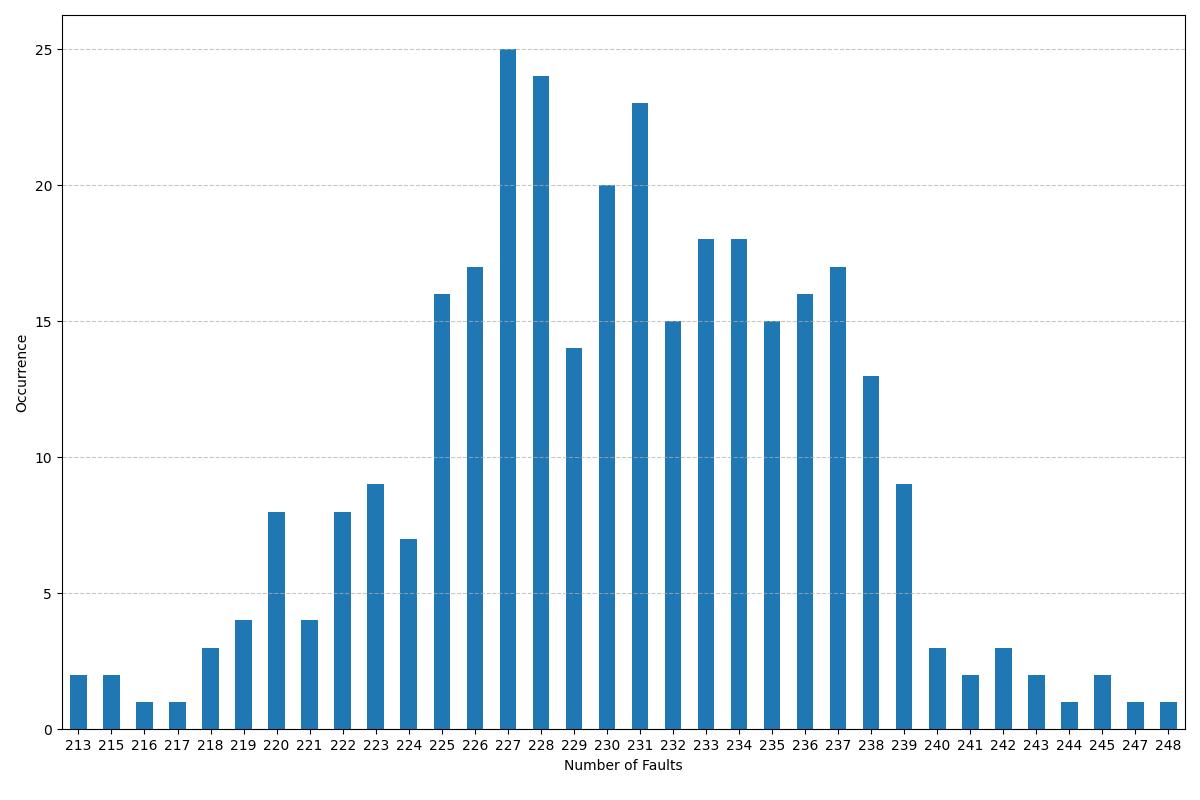}
    \caption{Frequency of the number of faults for each successful attack.}
    \label{morus exp result}
\end{figure}

In the remaining 324 cases, the results show that the full initial state of MORUSv2 can be successfully recovered. Among these 324 successful attacks, 86 cases directly recovered the state by solving equations, while 238 required the adversary to guess a few state bits in order to achieve full state recovery. For these 238 cases, the recovered state bits can be further categorized into three groups: directly recovered bits, indirectly recovered bits and guessing bits. The directly recovered bits refer to the state bits that can be directly recovered by solving the equations. The guessing bits are the state bits that cannot be recovered by solving the equations, and the adversary is required to guess the value to recover the state. The indirectly recovered bits are the state bits that can be recovered from the equations that involved one or more guessing bits. To recover the indirectly recovered bits correctly, the adversary needs to guess the value of the guessing bits correctly. In these attacks, the number of directly recovered bits ranged from 600 to 639, while the number of indirectly recovered bits and guessing bits ranged from 0 to 35 and 1 to 6, respectively. 
Considering the highest number of guessing bits of 6, the adversary only needs to try for $2^6$ possible combinations at most. With the correct guess, the indirectly recovered bits are resolved, leading to a full recovery of the initial state. The average number of faults required to recover the initial state is 231. Figure~\ref{morus exp result} illustrates the distribution of the number of faults required for a successful state recovery of MORUS across the conducted experiments.

The attack proposed by Wong et al. \cite{Wongetal2020} can recover only two register's contents under the assumption of multi-byte random fault with precise control over fault location. In contrast, our proposed attack utilizes a more relaxed assumption in terms of the fault precision (no control over the target register). 
Consequently, our proposed attack require a weaker fault model while achieving full initial state recovery of MORUSv2. Although Wong et al. \cite{Wongetal2020} performed a theoretical analysis, they did not implement the attack on the cipher. Hence, the proposed method is considered as the first implementation of differential fault attack against MORUS. 
The total complexity of the attack is bounded by the Gr\"obner basis computation for solving the equation system and guessing the values of $h$ guessing bits, where $h\in\{1,2,3,4,5,6\}$. Since the time complexity of Gr\"obner basis solver to resolve this specific system is negligible, the overall complexity of the attack is at most $2^h$.

\subsection{Differential Fault Attack on ATOM}
\label{imp dfa atom}
In the attack on ATOM, we assumed that the adversary knows 56 bits of plaintext $P=\{p_0,p_1,\cdots,p_{55}\}$. Due to the design of the cipher, the attack assumption made for this attack is more stringent than those in previous sections. Since the $cnt$ in the keystream generation phase for the key filter depends on the state of LFSR, the adversary needs to possess the information of the LFSR state to determine which key bits are involved in the operation. Hence, it is assumed that the adversary knows the first 60 bits of the LFSR. The adversary does not need to guess the value of the last 9 bits of the LFSR, as they are initialized to a constant value of 1 at the beginning of encryption phase. The fault is assumed to be injected at the initial state of NFSR.

\subsubsection{Implementation of MLP Model for Fault Location Identification}
The implementation of the MLP model for ATOM is focused on identifying the fault location in the NFSR. Thus, the generated datasets only include samples of faults injected in the NFSR. The training and hyperparameters tuning follow the same method as described earlier. The architecture of the developed MLP model is described as follows:
\begin{itemize}
\itemsep0em
    \item \textbf{Input layer:} The number of neurons in the input layer is 56. 
    \item \textbf{Hidden layer:} Three hidden layers, with 128, 256, 128 neurons respectively, are used. 
    For the first two hidden layers, the ELU activation function and the dropout rate of 0.2 are applied after the batch normalization to reduce overfitting. The last hidden layer utilizes the ELU activation after batch normalization. 
    \item \textbf{Output layer:} 90 neurons are used in this layer (number of possible fault locations in the NFSR of ATOM), using softmax as the activation function. 
\end{itemize}

\subsubsection{Implementation of Differential Fault Attack on ATOM}
We describe the implementation of DFA on ATOM in Algorithm~\ref{dfa atom}.  
In this attack, the threshold $\mathcal{T}$ is set to 234. This value accounts for the 90 differential equations required to determine the 90-bit NFSR state, 128 differential equations to determine the 128-bit secret key, and an additional 16 original fault-free equations included to further constrain the system.
Note that only a partial recovery of the secret key may be possible, as symbolic generation in SageMath was limited to the first 16 keystream equations. These constraints are insufficient to uniquely resolve all 128 secret key bits.
\begin{algorithm}[!b]
\caption{Differential Fault Attack on ATOM}
\label{dfa atom}
\begin{algorithmic}
    \STATE Initialize list $\mathcal{E}, \mathcal{F}$
    \STATE $solve \gets$\FALSE, $\mathcal{T}\gets234$, $num\_eq\gets0$
    \STATE $Z=\{z_0, z_1, \cdots,z_{55}\}\gets encATOM(S, P)$
    \STATE Generate fault-free ANF, $NKS=\{nks_0,\cdots,nks_{15}\}$
    \FOR{$i=0$ \textbf{to} 15}
        \STATE Add $nks_i\oplus z_i$ to $\mathcal{E}$
        \STATE $num\_eq = num\_eq + 1$
    \ENDFOR
    \WHILE{$solve$ \textbf{is} \FALSE}
        \WHILE{$num\_eq<\mathcal{T}$}
            \STATE Reset cipher to initial state $B^\prime, L$
            \STATE Inject fault at location $f$, where $b_f=b_f\oplus1$
            \STATE $Z^\prime =\{{z}_0^\prime, z_1^\prime, \cdots,z_{55}^\prime\}\gets encATOM(B^\prime, L)$
            \FOR{$i=0$ \textbf{to} 55}
                \STATE $\Delta z_i^f = z_i \oplus z_i^\prime$
            \ENDFOR
            \STATE $f^\prime\gets modelATOM(\Delta Z^f=\{\Delta z_0^f,\cdots,\Delta z_{55}^f\})$ 
            \IF{$f$ \textbf{not in} $\mathcal{F}$ \AND $f=f^\prime$}
            \STATE Append $f$ to $\mathcal{F}$
            \ELSE
            \STATE \textbf{continue}
            \ENDIF
            \STATE Generate faulty ANF, $FKS=\{fks_0,\cdots,fks_{15}\}$
            \FOR{$i=0$ \textbf{to} 15}
                \STATE $dks=nks_i\oplus fks_i\oplus\mathrm{\Delta}z_i^f$
                \IF{$dks.deg\ge1$ \AND $dks$ \textbf{not in} $\mathcal{T}$}
                \STATE Append $dks$ to $\mathcal{E}$
                \STATE $num\_eq = num\_eq + 1$
                \ENDIF
            \ENDFOR
        \ENDWHILE
        \STATE Solve $\mathcal{E}$ with Gr\"obner basis
        \IF{Gr\"obner basis can solve in time}
            \STATE $solve \gets$\TRUE
        \ELSE
            \STATE $\mathcal{T}=\mathcal{T}+5$
        \ENDIF
    \ENDWHILE
\end{algorithmic}
\end{algorithm}

In our experiments, we first generate and store the fault-free keystream equations into $\mathcal{E}$ by XOR-ing $nks_i$ with the corresponding keystream bit $z_i$, where $i=\{0,1,\cdots,15\}$.
Next, a bit-flipping fault is injected at a random location $f$ in the initial state of NFSR $B$, such that $b_f$ is complemented. Once the  differential keystream, $\Delta Z^f={\Delta z_0^f,\Delta z_1^f,\cdots,\Delta z_{55}^f}$ is obtained, it is fed into the MLP model to identify the fault location. If the fault location is correctly identified, the differential equations $dks$ are retrieved by $nks_i\oplus fks_i\oplus\Delta z_i^f$, where $i=\{0,1,\cdots,15\}$ and added to $\mathcal{E}$, regardless of their algebraic degree. The same process of generating and collecting differential equations is repeated until the number of equations in the equation system reaches or exceeds the threshold $\mathcal{T}$ . 
The system of equations is solved using Gröbner basis when the number of equations stored in $\mathcal{E}$ exceeds the set threshold. 
The solving phase follows the same methodology used for ACORNv3 and MORUSv2. In instances where the solver does not return a solution within the specified time limit, the threshold is increased by 5 to include more equations in the equation system.

\subsubsection{Result and Discussion of the Attack on ATOM}
Table~\ref{atom model performance} shows the performance of the MLP model for ATOM. It shows that the performance of the trained model achieves an accuracy of 82.3568\%. Compared to the trained models for ACORNv3 and MORUSv2, this model has a lower accuracy on ATOM. Despite achieving an accuracy exceeding 80\%, the model remained susceptible to misidentification errors when applied to ATOM. This is primarily due to the cipher's design structure. ATOM utilizes a double key filter where the setting of $cnt$ in the encryption phase depends on the current LFSR state. This dependency introduces a high degree of complexity in the fault propagation, making the resulting patterns less deterministic and harder for the MLP to learn. 
The model achieves a precision score of 82.1186\%, indicating that there is a moderate chance that it will encounter false positive identifications. The recall score of 82.3568\% is slightly higher than the precision score, showing that the probability of misidentification is slightly lower than the probability of false positive identifications. Finally, the F1 score of 81.6163\% shows a balanced trade-off between the precision and recall. 
\begin{table}[!htbp]
\centering
\caption{Performance of trained model for identifying fault locations on ATOM}
\label{atom model performance}
\begin{tabular}{cccc}
\hline
Accuracy & Precision & Recall & F1-score \\
\hline
0.823568 & 0.821186 & 0.823568 & 0.816163 \\
\hline
\end{tabular}
\end{table}

On the other hand, the traditional signature-based method is also implemented to benchmark the performance of the trained model. The same training and testing datasets are used to generate the fault signature and evaluate the signature-based method. The accuracy of the MLP model is 82.3568\%, while the accuracy for the signature-based method is 58.8976\%. The results show that the traditional signature-based method has a poor performance in fault location identification for ATOM. 
Overall, the MLP model demonstrates a significant performance improvement over the signature-based method in learning complex fault propagation patterns, exceeding its accuracy by a margin of approximately 23\%.
Although the performance of this model is not as significant as the previously trained models for ACORN and MORUS, it nonetheless demonstrates that deep learning model possesses a superior capability compared to traditional signature-based methods to learn the highly non-deterministic fault propagation patterns.  

A total of 400 experiments were conducted to evaluate the performance of the proposed state recovery and key recovery attack on ATOM. 
Under the strict parameters of this study, the attack only proceeds to the equation-solving phase if all fault locations are identified with 100\% accuracy. However, in all 400 trials, at least one misidentification occurred during the collection process.
These results show that the utilization of a double key filter provides a relatively stronger security margin against the attacks presented in this paper, as it may prevent the adversary from constructing a system of equations where every equation is valid. 
If the adversary utilizes the equation system with the misidentification cases and attempts to solve the system, it is expected that some of the state bits or the key bits will not be recovered correctly. In our experiments, a new fault is injected into the NFSR whenever the number of equations is below the threshold of 234. To reach this threshold, the attack requires an average of 46 faults. 

Since the attack under the random fault location model (where the adversary has no control over fault locations) encountered misidentification errors in all 400 trials, a subsequent experiment was conducted assuming a precise fault model. In this experiment, the utilization of threshold is excluded. 
Based on the previous experiments, 46 faults are needed on average to construct the equation system with at least 234 equations. Suppose that the adversary repeatedly reset the cipher and injects each fault in $\mathcal{F}$, where $\mathcal{F}$ consists of 46 faults that are defined as follows:
\begin{equation}
\begin{split}
    \mathcal{F}=\{&b_3,b_7,b_9,b_{10},b_{11},b_{16},b_{17},b_{20},b_{21},b_{22},b_{24},b_{25},b_{30}, b_{31}, b_{34},b_{35}, b_{36},b_{37},b_{40},b_{41}, \\& b_{42}, b_{44}, b_{46}, b_{48},b_{49}, b_{52},b_{53},b_{54},b_{56},b_{58},b_{62},b_{63},b_{65},b_{67},b_{69},b_{72},b_{73}, b_{75}, b_{76}, \\& b_{79},b_{82},b_{83},b_{84},b_{85},b_{86},b_{89}\}.
\end{split}
\end{equation}
The fault locations in $\mathcal{F}$ are chosen randomly, and subsequently the corresponding differential equations were generated, and included in the equation system together with the 16 fault-free equations. 
Because the secret key bits remain as unresolved variables within the differential equations, a brute-force guessing strategy was employed to facilitate the solving process. By iterating through the possible values of the involved key bits, the system becomes sufficiently constrained to allow the Gröbner basis solver to operate. Under these conditions, 74 out of the 90 NFSR initial state bits were successfully recovered.
However, some state bits were recovered with incorrect values; this is likely due to the limited keystream length, which leads to an underdetermined system where multiple mathematical solutions satisfy the given equations

\section{Conclusion}
\label{conclusion}
This work investigates the feasibility of deep learning-assisted differential fault attacks (DFA) against ACORNv3, MORUSv2, and ATOM, utilizing a single-bit bit-flipping fault with no control over the fault location. 
We implemented three Multilayer Perceptron (MLP) models to identify fault locations, each achieving higher accuracy than traditional signature-based methods.
Additionally, we proposed a threshold-based strategy to optimize the number of fault injections. Our results demonstrate that the initial state of ACORN can be recovered with negligible complexity, and full state recovery can be achieved for MORUS—both results outperforming the existing literature. 
Furthermore, we implemented the first DFA on ATOM, recovering the majority of the NFSR state bits under a precise fault-location model. These findings suggest that the double key filter design increases the security margin against the described fault attacks. Future work may examine the feasibility of deep learning-assisted fault attacks under a more relaxed random fault model, where the attacker has no control over the fault location or impact. 

\bibliography{reference.bib}





\end{document}